\documentclass[preprint,showpacs,preprintnumbers,amsmath,amssymb]{revtex4-2}

\usepackage{amsmath}
\usepackage{amsfonts}
\usepackage{amssymb}
\usepackage{graphicx}
\usepackage{color}
\usepackage{subcaption}
\usepackage{setspace}
\usepackage{csquotes}

\usepackage{hyperref}
\hypersetup{
    colorlinks=true,
    linkcolor=red,
    filecolor=magenta,      
    urlcolor=cyan,
    pdfpagemode=FullScreen,
    }

\usepackage{array}

\makeatletter
\newcommand{\thickhline}{%
    \noalign {\ifnum 0=`}\fi \hrule height 1pt
    \futurelet \reserved@a \@xhline
}
\newcolumntype{"}{@{\hskip\tabcolsep\vrule width 1pt\hskip\tabcolsep}}
\makeatother

\begin{document}

\title{Dependency of quantum time scales on symmetry}

\author{Fei Guo$^{1,2}$}
\author{Dmitrii Usanov$^{3}$}
\author{Eduardo B. Guedes$^{3}$}
\author{Mauro Fanciulli$^{4,5,6}$}
\author{Kaishu Kawaguchi$^{7}$}
\author{Ryo Mori$^{7}$}
\author{Takeshi Kondo$^{7,8}$}
\author{Arnaud Magrez$^{1}$}
\author{Michele Puppin$^{1,2}$}
\author{J. Hugo Dil$^{1,2,3}$}

\affiliation{
$^{1}$Institute of Physics, \'{E}cole Polytechnique F\'{e}d\'{e}rale de Lausanne, CH-1015 Lausanne, Switzerland\\ 
$^{2}$Lausanne Centre for Ultrafast Science (LACUS), \'{E}cole Polytechnique F\'{e}d\'{e}rale de Lausanne (EPFL),
CH-1015 Lausanne, Switzerland\\
$^{3}$Center for Photon Science, Paul Scherrer Institut, CH-5232 Villigen, Switzerland\\
$^{4}$CY Cergy Paris Universit\'{e}, CEA, LIDYL, 91191 Gif-sur-Yvette, France\\
$^{5}$Universit\'{e} Paris-Saclay, CEA, LIDYL, 91191 Gif-sur-Yvette, France\\
$^{6}$New Technologies Research Center, University of West Bohemia, 30100 Plzen, Czech Republic\\
$^{7}$Institute for Solid State Physics, The University of Tokyo, Kashiwa, Chiba 277-8581, Japan\\
$^{8}$Trans-scale Quantum Science Institute, The University of Tokyo, Tokyo 113-0033, Japan
}

\date{\today}
\begin{abstract}
Although used extensively in everyday life, time is one of the least understood quantities in physics, especially on the level of quantum mechanics. Here we use an experimental method based on spin- and angle-resolved photoemission spectroscopy from spin-degenerate dispersive states to determine the Eisenbud-Wigner-Smith (EWS) time delay of photoemission. This time scale of the quantum transition is measured for materials with different dimensionality and correlation strength. A direct link between the dimensionality, or rather the symmetry of the system, and the attosecond photoionisation time scale is found. The quasi 2-dimensional transition metal dichalcogenides 1T-TiSe$_2$ and 1T-TiTe$_2$ show time scales around 150 as, whereas in quasi 1-dimensional CuTe the photoionisation takes more than 200 as. This is in stark contrast with the 26 as found for 3-dimensional pure Cu. These results provide new insights into the role of symmetry in quantum time scales and may provide a route to understanding the role of time in quantum mechanics.

\end{abstract}

\maketitle
\section{Introduction}

The role of time in quantum mechanics remains one of the most fundamental problems in the field, despite close to a century of physical and philosophical endeavours. One direction towards resolving this conundrum is to probe the time scale of quantum processes and find interdependencies of time and other variables. For example, chronoscopies of tunneling ionization and photoionization probe attosecond timescales close to the atomic unit of time and have thereby established new boundaries for accessing fundamental time scales \cite{Pazuorek:2015}. In both of these processes, a bound electron is transferred to a free continuum state under the action of an electromagnetic field. In the case of tunneling ionization, intense fields bend the confining potential, allowing for tunneling through a barrier; whereas in photoionization, energetic photons directly drive a quantum transition to a continuum state.

Such phenomena have long been approximated as practically instantaneous, until the emergence of attosecond time-resolved spectroscopies falsified this oversimplification and reignited the fundamental problem of defining time in quantum mechanics \cite{Dahlstrom_2012, Pazuorek:2015}. As a natural follow-up, improving measurement precision of attosecond change of quantum state is not only beneficial for many scientific and technological applications, but also for fundamental research. Examples are improving quantum state control and read-out \cite{Haroche:2013}, enhancing the sensitivity of relativity tests \cite{Cacciapuoti:2009}, and providing a sensitive probe for quantum many-body effects \cite{Zhang:2014}. As discussed in more detail below, recent studies have explored alternative routes to direct attosecond-resolved techniques, showing that, fundamental time scales
can even be determined without any external temporal reference, i.e. without a clock.

In the case of quantum tunneling, different methods have been proposed to measure the time a particle spends inside the classically forbidden region. By comparing the center-of-mass positions of a tunneled particle and a free particle with the same momentum, the so-called Wigner phase time can be calculated using the group velocity of a wave packet \cite{Buttiker:1983, Hauge:1989}. However, since this time diverges for small tunneling probabilities, superluminal group velocities have been calculated from experimental results \cite{Steinberg:1993, Spielmann:1994}. Alternatively, in the strong-field ionization regime, a time-varying potential barrier induced by elliptically polarized radiation gives rise to a time-dependent scattering angle, from which tunneling times can be calculated \cite{Uiberacker:2007, Eckle:2008, Satya:2020}. This attoclock approach has revealed instantaneous tunneling in some cases \cite{Eckle:2008}. However, the calculation of tunneling time depends heavily on the ionization model used \cite{Sainadh:2019, Satya:2020}.

An operational definition of the travel time within the quantum barrier can be based on a measurement using a Larmor clock \cite{Buttiker:1983, Ramos:2020}. By spatially overlapping the barrier with a magnetic field that causes spin precession, the dwell time within the barrier can be linked to the angle of precession and the Larmor frequency. Recently, along the lines of Ramsey interferometry \cite{Nicholson:2015}, a novel approach of accessing the tunneling time has been proposed \cite{Schach:2024}, which starts with a coherent superposition of atoms with 2 internal states $|g\rangle$ and $|e\rangle$. After adding a quantum barrier in the non-interacting region of a conventional Ramsey interferometer, the interference signal read out contains not only information about the laboratory time $t$ during which the atom travels within the Ramsey clock plus relativistic time dilation $\delta t$, but also information about the tunneling time $\tau$. Namely, the Ramsey clock induces interference and measures the phase difference between the 2 levels after tunneling $|g_T\rangle$ and $|e_T\rangle$: $arg \langle e_T | g_T \rangle = \Delta\omega(t+\delta t+\tau)$, in which $\Delta \omega=\omega_e - \omega_g$ is the transition frequency. While the contribution from $t+\delta t$ comes from the evolution of the states themselves, the contribution from the tunneling time $\tau$ comes purely from the complex part of the tunneling amplitude $T_{e/g}$, and can be expressed as a function of the atom's kinetic energy and the barrier profile. Finally, this phase difference due to $\tau$ can be isolated by introducing a second, counter-propagating Ramsey clock with no quantum barrier as a reference, which cancels out the $t+\delta t$ phase contribution. While currently such a Ramsey clock has not been experimentally realized yet, similar physics is readily accessible in photoionization.

It comes as an advantage that photoionization has well-understood energetics and kinematics and is widely used in spectroscopy to explore the electronic structure of atoms, molecules, and solids \cite{Damascelli:2004}. Since the 1950s, theory has postulated that photoemission can be seen as a half-scattering process that induces a phase shift in the outgoing wave packet \cite{Wigner:1955,Eisenbud:1948,Smith:1960} relative to a wave packet propagating in vacuum \cite{Nussenzveig:1972}. In the Eisenbud-Wigner-Smith (EWS) model, the half-scattering time delay is related to the phase shift of the scattered wave packet by:
\begin{equation}
    \tau_{\text{EWS}}=\hbar\frac{d\phi}{d E_k},
    \label{EWS}
\end{equation}
where \( \phi \) is the phase shift, and \( E_k \) is the kinetic energy of the outgoing photoelectron.

Attosecond chronoscopy of the photoemission process \cite{Eckle:2008,Schultze:2010,Klunder:2011} has revealed relative delays in the attosecond range between electrons emitted from different states in the same material, and it can thus determine the \emph{relative} EWS time, using as a reference a different electronic state within the same system \cite{Cavalieri:2007,Neppl:2012, Lucchini:2015} or a second reference system \cite{Locher:2015}. In order to access the \emph{absolute} EWS time, in analogy to the tunneling amplitude $T_{e/g}$, the phase term \( \phi \) of the transition matrix element \( M_{fi}= \left\langle \psi_f \middle| \hat{H}_{int} \middle| \psi_i \right\rangle = R e^{i\phi} \) is required. Phase information is typically lost when probing photoemission intensity; thus, one needs to search for other physical observables that contain explicit information about the phase of the interaction matrix element.

One example of such observables is the angular distribution of photoelectrons from atomic or molecular photoionization \cite{Wang:2001, Hockett:2016}, and it has been demonstrated that the phase term can be extracted by circular dichroism in UV photoelectron diffraction experiments \cite{Daimon:1995b, Wiessner:2014}. However, this method is not practical for crystalline solids, since the small changes in angular distribution are overshadowed by the intrinsic energy-momentum distribution relations of the band structure.

The spin of photoelectrons emitted at a given angle also depends explicitly on the matrix element phase \cite{Huang:1980, Kessler:1985}. Although initial studies mainly focused on photoionization from atoms in the gas phase, in the case of solids, spin-degenerate initial states result in a spin polarization, which depends on the matrix element phase \cite{Heinzmann:2012}. Recently, we developed a semi-analytical model to estimate the EWS photoemission time delay from spin polarization of photoelectrons emitted from solids \cite{Fanciulli:2018}. This technique, based on spin- and angle-resolved photoemission spectroscopy (SARPES) \cite{Dil:2009R}, has been used in estimating the lower boundary of the EWS time delay \( |\tau_{EWS}|\geq26 \, \text{as} \) from a Cu(111) single crystal \cite{Fanciulli:2017}. However, to obtain insight in the role of time in quantum mechanics, it is essential to determine what factors influence such fundamental time scales.
%In addition to obtaining such fundamental time scales, the phase sensitivity of EWS delays makes them a promising probe of microscopic correlations and of many-body interactions. 

\begin{figure}
    \centering
    \includegraphics[width=0.9\linewidth]{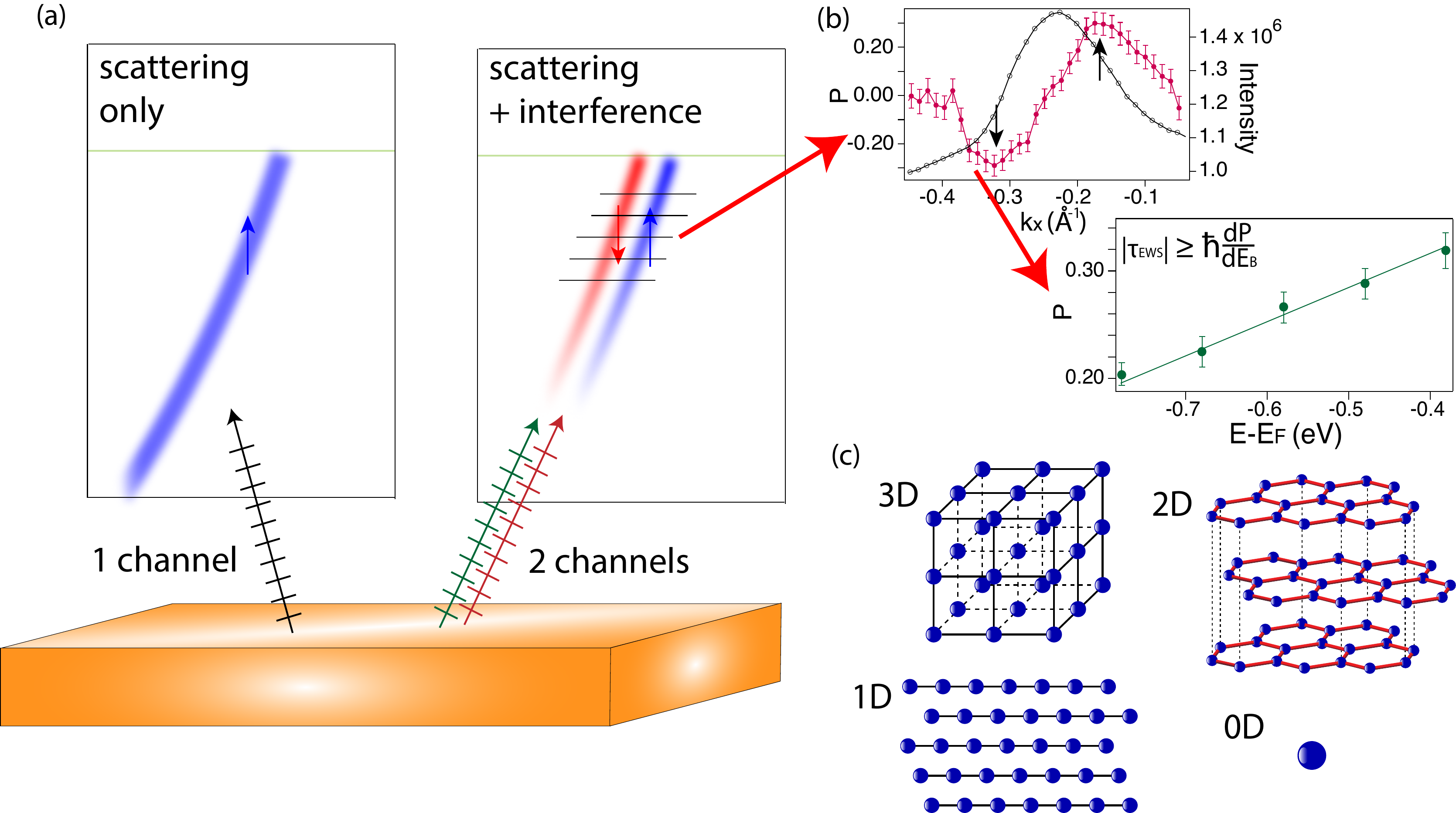}
    \caption{(a) Schematic drawing of 1-channel and 2-channel photoemission, along with the subsequent single and double polarization feature in spin polarization, (b) example of measured $P(k)$, and $P(E)$ constructed from $P(k)$ at various energies, where the slope is used to calculate $|\tau_{EWS}|$, (c) schematic drawing of matter in 3, 2, 1 and 0 dimensions.}
    \label{schematicsD}
\end{figure}

In this light, it is interesting to note that the EWS photoemission time delay in the high-temperature superconductor Bi$_2$Sr$_2$CaCu$_2$O$_{8+\delta}$ (BSCCO) was determined as \( |\tau_{EWS}|\geq120 \, \text{as} \) \cite{Fanciulli:2017B}, much larger than that of Cu(111). It is tempting to link this delay to stronger electronic correlations in BSCCO compared to Cu(111).
%yet, the way of quantifying correlation strength is not uniquely defined, and may not be perfectly reliable for crystalline solids (see ref. \cite{Izsak:2023} and references therein). 
However, a distinctive and measurable difference lies in the dimensionality of the crystals and their electronic structure, e.g., BSCCO being quasi-2-dimensional (2D) and Cu being 3-dimensional (3D). Fundamentally, crystal symmetry is embedded in dimensionality, as shown schematically in Fig.(\ref{schematicsD})(c): an atom is 0D and highly symmetric, and further the level of symmetry reduces from 3D to 1D. In this work, using dimensionality as a simple way of quantifying the degree of symmetry, we extend the study to materials of various dimensionalities and exhibiting different broken symmetry ground states. We focus on two quasi-2D transition metal dichalcogenides (TMDCs): 1T-titanium diselenide (1T-TiSe$_2$) with charge-density-wave order, 1T-titanium ditelluride (1T-TiTe$_2$) without charge-density-wave order; and a quasi-1D material copper telluride (CuTe) with charge-density-wave order. We find the general trend that $\tau_{EWS}$ increases with decreasing dimensionality, and thus generally with reduced symmetry.

\section{Results}
\subsection{Model to extract EWS time delay}

The EWS time delay in photoionization is given by:
\begin{equation}
    \tau_{EWS}=\hbar\frac{d\phi}{dE}=\hbar\frac{d\angle\left(\left\langle \psi_f \middle| \hat{H}_{int} \middle| \psi_i \right\rangle\right)}{dE}.
\end{equation}
where $\angle$ draws out the complex phase of the matrix element. 
In order to make $\tau_{EWS}$ accessible with SARPES, the crucial step is to bridge between $\phi=\angle\left(\left\langle \psi_f \middle| \hat{H}_{int} \middle| \psi_i \right\rangle\right)$ and the photoelectron spin. In the picture of electron scattering, because of the spin-orbit coupling term in the scattering potential, the differential scattering cross section is spin dependent, and an unpolarized electron beam can become spin polarized upon scattering \cite{Kessler:1985}. The same argument is valid in the case of photoemission, an initially spin degenerate state can produce spin polarized photoelectrons even with unpolarized or linearly polarized light that do not carry spin angular momentum \cite{Cherepkov:1983}, and the spin polarization was shown to be a result of spin-orbit-induced hybridization of different basis functions representing different single-group spatial symmetries \cite{Tamura:1991b, Irmer:1992}. 

To illustrate this concept, we consider the simplest case. When linearly polarized radiation hits the crystal with an off-normal incidence angle, its electric field can be decomposed into $E_{\parallel}$ and $E_{\perp}$ which are parallel and perpendicular to the crystal surface, and each of them gives rise to corresponding complex transition matrix elements $M_1$ and $M_2$ respectively. The resulting magnitude of spin polarization of photoelectrons $\textit{\textbf{P}}=\sum_{i=x,y,z}\hat{\mathbf{i}}\frac{N^{\uparrow}_i-N^{\downarrow}_i}{N^{\uparrow}_i+N^{\downarrow}_i}$ is given by the interference, or relative phase difference, of these 2 photoemission channels $P\propto \Im[M_1 M^*_2]$. With the magnitude of \textit{\textbf{P}} now expressed as a function of the phase shift, one can subsequently proceed to estimate $\tau_{EWS}$ and obtain $|\tau_{EWS}|\geq\hbar|\frac{dP}{dE_k}|$ \cite{Fanciulli:2018}. The direction of \textit{\textbf{P}} is perpendicular to the reaction plane determined by the  incident light and the symmetry of the state under investigation\cite{Tamura:1991b}. This can be used to further refine the obtained time scale in terms of total and scattering time \cite{Fanciulli:2018}. However, this requires assumptions that might not hold between different experimental geometries and material systems, and will thus not be employed here.

To resume, it is the intrinsic interference of different photoemission channels that gives rise to the observable \textit{\textbf{P}}, which is analogous to the measured interference signal of the 2 tunneling levels $arg \langle e_T | g_T \rangle$ in the case of the tunneling Ramsey clock \cite{Schach:2024}. As elaborated in the methods section and illustrated in Fig.(\ref{schematicsD})(a), the presence of a Feshbach-type resonance and the consecutive reversal of the spin direction across the band maximum, can be used to confirm this intrinsic multi-channel interference. We refer to this as double polarization feature (DPF) in the measured spin polarization. A similar sign-reversal feature has been observed in the dichroism asymmetry in circular dichroism ARPES, where it is also considered a result from interference effects \cite{Schmitt:2024arXiv}.

Here, we use the above model to estimate the photoemission time scale from crystalline solids of different dimensionality. With linearly polarized quasi CW light incident at off-normal angles, we consider spin polarization of photoelectrons from spin-degenerate initial states to arise from interference between the most prominent photoemission channels, which are generally non-zero and assumed to be of the same order of magnitude. With a clear signature of multi-channel interference as shown schematically in Fig.(\ref{schematicsD})(a), we measure with random order the spin polarization of photoelectrons at various different binding energies, construct $P(E)$ and extract $|\tau_{EWS}|\geq\hbar|\frac{dP}{dE_k}|$, as shown in Fig.(\ref{schematicsD})(b).

\subsection{Quasi-2D 1T-TiSe$_2$ and 1T-TiTe$_2$}
%Low-dimensional materials are attracting continued interest due to their rich number of competing broken symmetry ground states \cite{Young:2014,Hedge:2022}, and one of the most significant phenomenon is the susceptibility to charge density ordering, such as charge density waves (CDWs). TMDCs have been a model system of study in this aspect \cite{Wilson:2001, Rossnagel:2011, Hwang:2024} as they stabilize in various structural polymorphs depending on the stacking symmetry, and changes in polymorph or element may imply a completely different CDW mechanism \cite{Rossnagel:2011, Yang:2017}.

%1T-TiSe$_2$ is one of the most studied CDW compounds among all TMDCs \cite{Suzuki:1984, Anderson:1985, Kidd:2002, Li:2007, Chen:2015, Chen:2016, Sugawara:2016, Hildebrand:2016, Singh:2017} due to its unconventional CDW mechanism: an ongoing debate between an excitonic insulator \cite{Jerome:1967, Cercellier:2007, Monney:2011, Monney:2015}, and the band Jahn-Teller effect \cite{Hughes:1977, Whangbo:1992, Rossnagel:2002}. Extensive experimental studies with spin-integrated ARPES and theoretical studies have been performed on 1T-TiSe$_2$ to investigate its electronic structure in order to understand the origin of its CDW phase \cite{Monney:2010, Vydrova:2015, Bianco:2015, Pasquier:2018, Watson:2019}.

To disentangle the possible influence of correlations from dimensionality, we compare the two nearly identical quasi-2D systems 1T-TiSe$_2$ and 1T-TiTe$_2$. Whereas the former is well known for its charge density wave \cite{Cercellier:2007, Monney:2015}, the latter does not show such correlation effects in its bulk form \cite{Claessen:1996, Rossnagel:2001}. Both have been investigated with ARPES \cite{Watson:2019, Strocov:2006}, but to our knowledge, there has  been no studies of the electronic band structure with spin resolution, and certainly no investigation of the photoemission time scale.

\begin{figure}
    \centering
    \includegraphics[width=\linewidth]{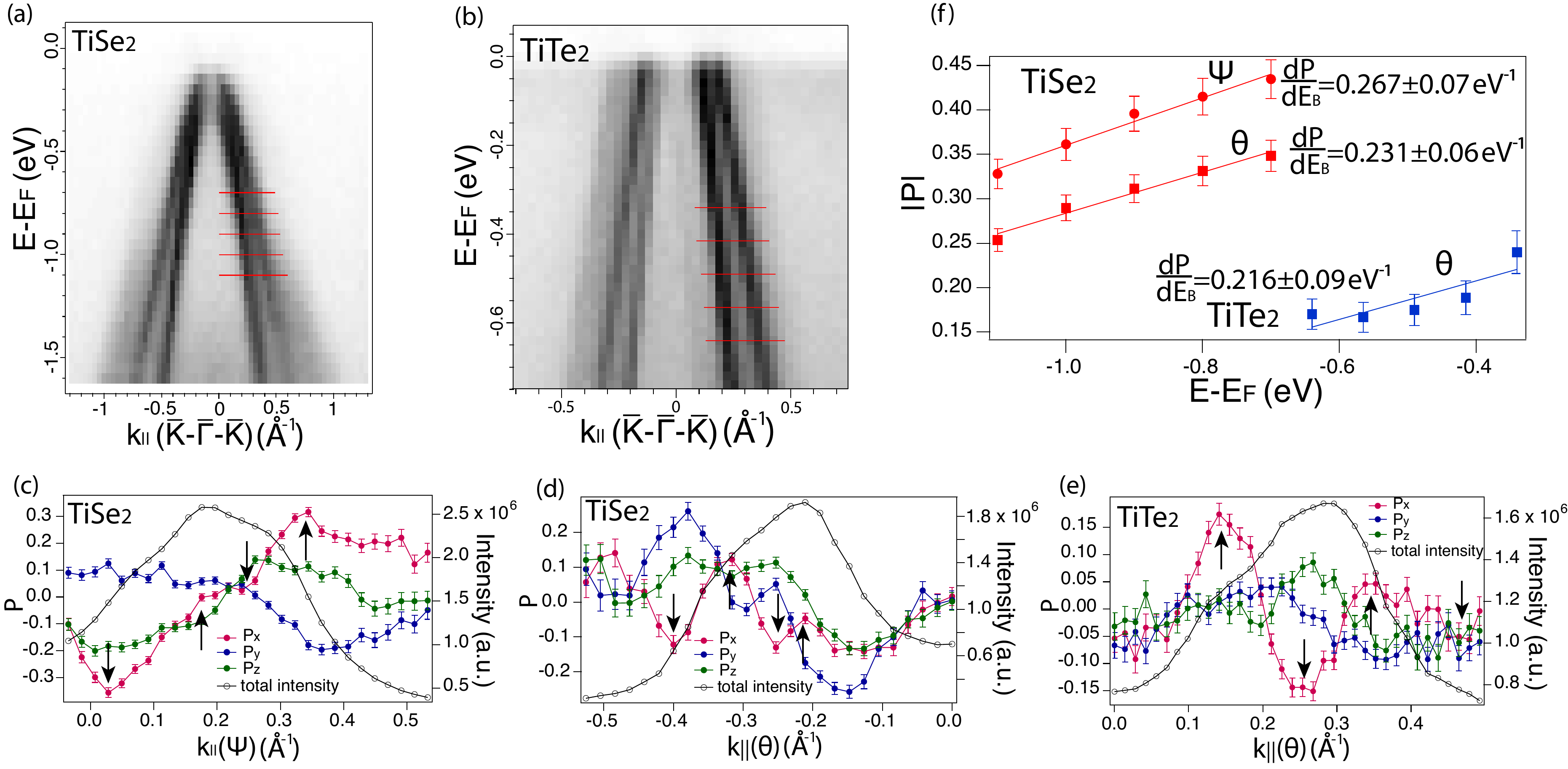}
    \caption{Band maps of (a) 1T-TiSe$_2$ and (b) 1T-TiTe$_2$ in the $\overline{K}-\overline{\Gamma}-\overline{K}$ direction, taken with $h\nu=67\,$eV, red lines indicate the binding energies at which momentum distribution curves (MDCs) were taken. MDCs of spin polarization of 1T-TiSe$_2$ at $E-E_F=-0.9\,$eV in x, y and z directions obtained by rotating the angle $\psi$ (c) and $\theta$ (d), together with total photoemission intensity. (e) MDCs of spin polarization of 1T-TiTe$_2$ at $E-E_F=-0.49\,$eV in x, y and z directions obtained by rotating the angle $\theta$. Black arrows in (c), (d) and (e) indicate 2 DPFs. (f) Maximum spin polarization magnitude at 5 binding energies extracted from MDCs, resulting in $|\tau_{EWS}|\geq 176\,$as from $\psi$-MDCs of 1T-TiSe$_2$, $|\tau_{EWS}|\geq 152\,$as from $\theta$-MDCs of 1T-TiSe$_2$, and $|\tau_{EWS}|\geq 142\,$as from $\theta$-MDCs of 1T-TiTe$_2$}
    \label{TiSe2}
\end{figure}

Here, we present an investigation of the spin polarization of the photoelectrons from the Se-$4p_{x,y}$ and Te-$5p_{x,y}$ derived valence bands. These show similar band structures along the $\overline{K}-\overline{\Gamma}-\overline{K}$ direction as illustrated in Fig.(\ref{TiSe2})(a) and (b). Due to inversion symmetry, these bulk bands are expected to be spin-degenerate. Considering first TiSe$_2$, on these bands, 5 indicated binding energies ranging from $E-E_F=-0.70$ to $-1.10\,$eV were chosen at which momentum distribution curves (MDCs) were taken for spin polarization in all spatial directions $x$, $y$ and $z$. These MDCs were taken by respectively scanning the tilt $\psi$ (Fig.(\ref{TiSe2})(c)) and polar $\theta$ Fig.(\ref{TiSe2})(d)) angles, i.e. by rotating around the sample $x$- and $y$-axes. Both measurements are along the $\overline{\Gamma}-\overline{K}$ direction in the Brillouin zone, and the sample azimuth was thus rotated by 30$^\circ$ in between.

In the MDCs of \textit{\textbf{P}} taken at $E-E_F=-0.9$\,eV in Figs.\ref{TiSe2}(c) and (d) all 3 components show clear momentum resolved spin polarization with a maximum value that exceeds $\pm 30\,\%$. From the total intensity and spin polarization, 2 overlapping double polarization features (DPFs) can be identified, as indicated by arrows in the respective figures. This confirms the presence of multichannel interference and we can thus proceed to extract the photoemission time scale.

The maximum magnitude of \textit{\textbf{P}} was calculated from all 3 spatial components for the peak around $k=0.33\,\mathring{\text{A}}^{-1}$ in Fig.\ref{TiSe2}(c), and from the peak around $k=-0.4\,\mathring{\text{A}}^{-1}$ in Fig.\ref{TiSe2}(d), summarized together in red in Fig.\ref{TiSe2}(f). Linear fits yield values of $\frac{dP}{dE}$ of $0.267\pm0.07\,$eV$^{-1}$ from $\psi$-MDCs and $0.231\pm0.06\,$eV$^{-1}$ from $\theta$-MDCs respectively. The corresponding minimum estimates for EWS time delays are $\tau_{EWS}\geq1.76\times 10^{-16}\,$s\,$ = 176\,$as and $\tau_{EWS}\geq1.52\times 10^{-16}\,$s\,$=152\,$as. This similarity in the magnitudes of spin polarization, as well as $\frac{dP}{dE}$ obtained by the 2 measurement geometries indicates that the value of $\tau_{EWS}$ we extract is not greatly affected by the experimental geometry. However, the spin polarization has noticeably different structures for the different angular directions. This indicates that the spin orientation is different due to the different experimental geometries and the subsequent different orbital contributions.

Similar to 1T-TiSe$_2$ we measured for 1T-TiTe$_2$\ $\theta$-MDCs at 5 binding energies ranging from $E-E_F=-0.34\,$eV to $-0.64\,$eV, as indicated on the band map in Fig.\ref{TiSe2}(b). MDCs of \textit{\textbf{P}} in all spatial directions taken at $E-E_F=-0.49eV$ are shown in Fig.\ref{TiSe2}(e). Again, spin polarization can be identified in all 3 directions, with the x-polarization particularly pronounced, showing 2 overlapping DPFs, as indicated by arrows. Using the same method as discussed above, for the peak around $k=0.14\,\mathring{\text{A}}^{-1}$, we obtain $P(E)$ as shown in blue in Fig.\ref{TiSe2}(f) with corresponding $\frac{dP}{dE}\approx0.216\pm0.09\text{eV}^{-1}$ and an estimate of of $\tau_{EWS}\geq1.42\times 10^{-16}\,$s$=142\,$as.

The found $\tau_{EWS}$ for 1T-TiTe$_2$ is between that obtained for 1T-TiSe$_2$ and BSCCO despite the less strong electronic correlation in the case of 1T-TiTe$_2$. Additionally, despite more complex interference effects, a time scale of $\tau_{EWS}>160\,$as has been extracted from spin polarization of the valence band of H-intercalated graphene \cite{Fanciulli:thesis}, which is clearly 2D. These results are strong indications that dimensionality or symmetry, rather than correlation, plays an important role in the magnitude of $\tau_{EWS}$. To further verify this hypothesis, we turn to quasi-1D CuTe, which, as typically in this dimension, exhibits CDW order.

\subsection{Quasi-1D CuTe}
Being quasi-1D and thus susceptible to charge ordering, CuTe has recently been demonstrated to host 3-dimensional CDWs \cite{Nguyen:2024, Guo:2024}. The electronic structure of CuTe has been studied with ARPES and time-resolved ARPES, and the quasi-1D states extending along the $\Gamma$-Y direction is observed to open a gap below $T_{CDW}=335\,$K \cite{Zhang:2018, Zhong:2024, Guo:2024}. However, the spin polarization of photoelectrons from these bulk-derived quasi-1D states has not been investigated.

\begin{figure}
    \centering
    \includegraphics[width=0.75\linewidth]{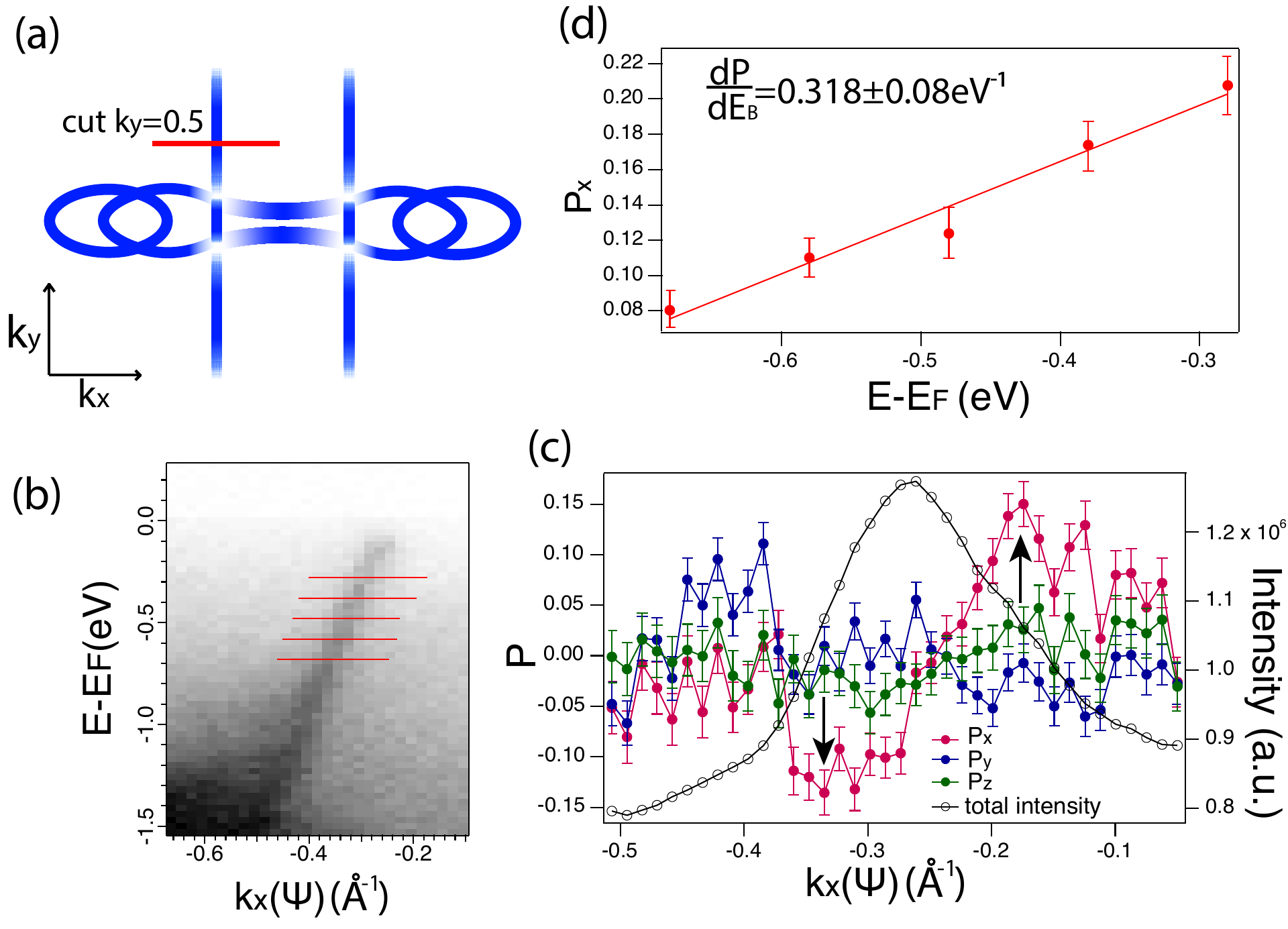}
    \caption{(a) Schematic drawing of the Fermi surface of CuTe with indication of the constant $k_y=0.5$ cut of the band map. (b) Band map of the 1D band taken at $k_y=0.5$ and $h\nu=26\,$eV, red lines indicate the binding energies at which MDCs were taken. (c) MDCs of spin polarization at $E-E_F=-0.48\,$eV in $x$, $y$ and $z$ directions obtained by rotating the angle $\psi$. Black arrows indicate the DPF. (d) Maximum spin polarization magnitude with each binding energy extracted from $\psi$-MDCs, resulting in $|\tau_{EWS}|\geq 209\,$as.}
    \label{CuTe}
\end{figure}

Fig.\ref{CuTe}(a) shows a schematic Fermi surface of CuTe, with the constant $k_y$ cut through the quasi-1D band indicated. Fig.\ref{CuTe}(b) shows the corresponding band map, with 5 binding energies ranging from $E-E_F=-0.28\,$eV to $-0.68\,$eV indicated where  $\psi$-MDCs of spin polarization are taken. As visible from the MDC at $E-E_F=0.48\,$eV in Fig.\ref{CuTe}(c), spin polarization is most pronounced along the $x$-direction reaching $\pm 15\,\%$, on which an anti-symmetric DPF is observed, whereas for the other 2 directions the measured spin polarization is very small. Considering \textit{\textbf{P}} in the y and z-directions as negligible, the $P_x$ magnitude of the peak around $k_x=-0.18\,\mathring{\text{A}}^{-1}$ in Fig.\ref{CuTe}(c) is plotted for 5 binding energies in Fig.\ref{CuTe}(d). From this we extract a value of $\frac{dP}{dE}\approx0.318\pm0.08\,$eV$^{-1}$ and a corresponding lower estimate of the EWS time delay $\tau_{EWS}>2.094\times 10^{-16}\,$s=209\,as. This time scale is significantly larger than that of all other materials investigated thus far, and it confirms the proposed relationship between EWS time scale and dimensionality, or more generally with reduced symmetry.

\section{Discussion}

The impact of spatial asymmetry on the EWS time delay has previously been considered by photoionization from gas phase Helium with a shake-up ionic final state \cite{Ossiander:2017}. For photons with sufficiently high energies, after photoemitting one electron to the ionization continuum, the electron remaining bound to the ion gets excited to one of the ionic Rydberg states, which have much larger spatial extent compared to the deeply bound ground state. This property of the shake-up state makes it strongly polarizable by the laser field inducing photoionization. During the attosecond time scale of photoionization, the He$^{+}$ ion left behind possesses a finite dipole moment and thus exerts a force on the outgoing photoelectron and causes a retardation. For the ionic Rydberg state $n=2$, this `correlation time delay' was calculated to be around 6\,as. In this context of photoionization, `correlation' is directly proportional to the effective dipole moment of the $He^{+}$ ion left behind, or more generally the extent of asymmetry, and directly affects the total EWS time delay. 

In the context of photoemission from solids, however, the degree of symmetry is manifested in crystal dimensionality as shown in Fig.(\ref{schematicsD})(c), with a decreasing number of mirror planes from 3D to 1D crystals. In this picture, the 0D atom has the highest symmetry, and one would expect a shorter $\tau_{EWS}$ than all the solid state materials investigated so far, which is indeed the case as seen by streaking experiments \cite{Schultze:2010, Eckle:2008}. Hence, our finding that $\tau_{EWS}$ is larger for crystals with reduced dimensionality corroborates with these previous studies and directly shows that reduced symmetry has a significant impact on the ionization time scale as summarized in Table.\ref{summary:time}. Interestingly, the $\tau_{EWS}$ obtained from the (quasi-)2D materials appears to directly reflect the expected interlayer coupling in these systems. Further studies are needed to show whether $\tau_{EWS}$ can indeed be used to quantify the transition in dimensionality.

\begin{table}[ht!]
\centering
\begin{tabular}{||c c c||} 
 \hline
 material & dimensionality & $\tau_{EWS}$(as) \\ [0.5ex] 
 \hline\hline
 Cu(111) \cite{Fanciulli:2017} & 3 & 26 \\ 
 \thickhline
 BSCCO \cite{Fanciulli:2017B} & 2 & 120 \\
 \hline
 TiTe$_2$ & 2 & 142 \\
 \hline
 TiSe$_2$ & 2 & 152 \\
 \hline
 graphene \cite{Fanciulli:thesis} & 2 & 160 \\
 \thickhline
 CuTe & 1 & 209 \\ [1ex] 
 \hline
\end{tabular}
\caption{ Summary of investigated materials with their lower estimated limit of $\tau_{EWS}$.}
\label{summary:time}
\end{table}

To summarize, we determined the absolute photoemission time scale for various materials using a model based on the measured spin polarization of photoelectrons.  A clear trend of increasing time scale with decreasing dimensionality is found, indicating a link between reduced symmetry and the quantum mechanical ionization time scale. For a detailed determination of the influence of correlation effects on the photoemission time scale it will therefore be essential to take the dimensionality of the states into consideration, along with other factors such as atomic species and orbital character. Under these conditions, spin-resolved attosecond chronoscopy has the potential to become a tool to characterize the nature and strength of interactions in correlated materials. 

Besides yielding fundamental information for understanding what determines the time delay in photoemission, complementary to attosecond streaking and RABBITT experiments, our results provide further insight into what factors influence time on the quantum level and might help understand the role of time in quantum mechanics. It will for example be of interest to investigate whether reduced symmetry has a general influence on coherence times in quantum operations. Furthermore, an increase of the duration of a quantum transition can generate additional degrees of freedom for quantum manipulations such as superposition of states or braiding.

\section{Materials and Methods}
1T-TiTe$_2$ and 1T-TiSe$_2$ single crystals are grown using the Chemical Vapor Transport (CVT) method with iodine (I$_2$) as the transport agent. High-purity Ti and Te or Ti and Se powders are sealed in a quartz ampoule and placed in a two-zone furnace with a controlled temperature gradient. For TiTe$_2$, the source is maintained at 800Â°C and the sink at 625Â°C, while for TiSe$_2$, the temperatures are slightly lower, with the source at 750Â°C and the sink at 600Â°C. After a few weeks, millimeter-sized crystals are obtained.

CuTe single crystals are grown using the self-flux method, where high-purity Cu and Te powders are mixed in a non-stoichiometric ratio with an excess of Te as the flux. The mixture is sealed in a quartz ampoule and heated to 600Â°C to form a homogeneous molten solution. The furnace is then slowly cooled at a controlled rate of 10Â°C per hour, allowing CuTe crystals to precipitate from the Te-rich melt. After cooling to room temperature, the excess Te flux is removed by sublimation.

SARPES measurements were taken at the COPHEE endstation on the SIS beamline of the Swiss Light Source, Paul Sherrer Institut. Band maps and spin polarization MDCs were measured with $\pi$-polarized light. Samples were prepared by attaching them to standardized sample holders and attaching a ceramic top post with conductive silver epoxy. Samples were cleaved at ultrahigh vacuum at T=25K, and the surface quality was checked with Low Energy Electron Diffraction (LEED). The measurement geometry is the same as that in ref.\cite{Fanciulli:2017}, in which the angle $\theta$ is scanned by rotating around the sample y-axis, and the angle $\psi$ is scanned by rotating around the sample x-axis. Spin asymmetry $A=\frac{N_{\uparrow}-N_{\downarrow}}{N_{\uparrow}+N_{\downarrow}}$ was measured by 2 orthogonally positioned classical Mott detectors. Spin polarization was calculated by $P_i=\frac{1}{S}A_i$ where the Sherman function $S=0.08$. MDCs were measured for different binding energies in a randomized order to exclude sample aging effects. The x, y and z components of spin polarization mentioned in text are in the sample frame. The zero polarization on all MDCs were calibrated with respect to the general anti-symmetry of the spin texture, and the magnitude of spin polarization was calculated for a single peak for each material.
All slopes and time scales are indicated in 3 (or less) significant digits, whereby the error is of the order of the second-to-last digit.

Double polarization features (DPFs) of spin polarization were observed in all materials investigated on their initially spin-degenerate bands. This was also seen on dispersive bands of Cu(111), BSCCO and graphene measured with synchrotron radiation, and was present on 1-step photoemission calculation of Cu(111) conduction band, but absent on core levels of Cu, and on dispersive bands of BSCCO measured with $h\nu=6.994\,$eV \cite{Fanciulli:2017, Fanciulli:2017B, Fanciulli:thesis}.

\begin{figure}
    \centering
    \includegraphics[width=0.95\linewidth]{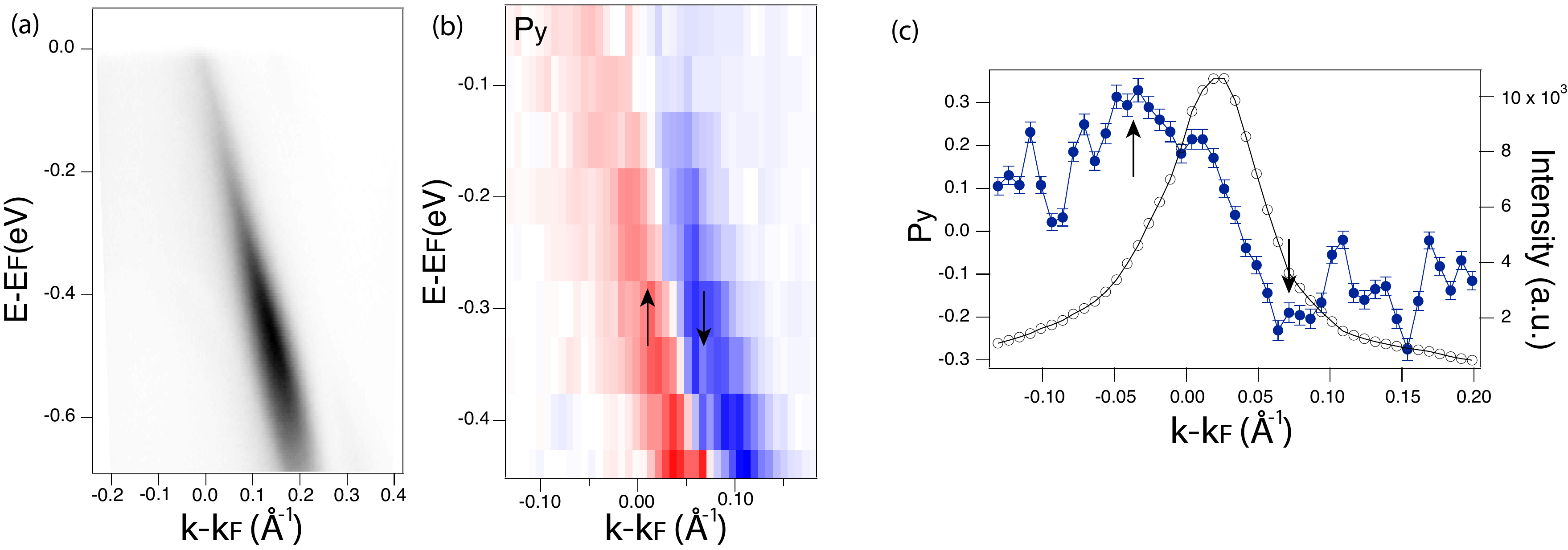}
    \caption{(a) Band dispersion of CuTe with $h\nu=6.994\,$eV. (b) spin polarization multiplied by the total intensity with quantization axis y. (c) y-spin polarization MDC obtained by moving electron deflector at $E-E_F=-0.2\,$eV. Black arrows indicate 2 opposite peaks in $P_y$.}
    \label{7eV}
\end{figure}

In order to rule out the possibility that this is a measurement artefact, further spin polarization measurements were performed on a 3D dispersive band of CuTe with $h\nu =6.994\,$eV at the laser ARPES facility of ISSP, University of Tokyo. The corresponding band map is shown in Fig.\ref{7eV}(a). With the experimental geometry used, this band shows mainly spin polarization in the y-direction, with corresponding polarization multiplied by total intensity shown in Fig.\ref{7eV}(b), and MDC of $\text{P}_y$ at $E-E_F=-0.2\,$eV shown in Fig.\ref{7eV}(c). A sign reversal can clearly be identified across the intensity maximum. This result further confirms the universality of a DPF in spin-degenerate bands. Below we explain how the presence of a DPF can be used to verify the existence of 2 or more interfering channels and thus that the model to extract $\tau_{EWS}$ can indeed be applied.

The DPF can be understood in close analogy to the Feshbach resonance, a well-studied effect in the field of ultracold atomic gases \cite{Fu:2014, Kurkcuoglu:2016, Takasu:2017}. A Feshbach resonance occurs when the energy of a scattering state in the continuum approaches that of a bound state, given that the bound state is coupled to the continuum. This effect is typically described with a 2-channel model: one channel represents the bound state and is referred to as`closed'; the other represents the continuum, and is referred to as `open'. In the fully elastic regime, the phase shift $\delta$ of the continuum wavefunction follows a Breit-Wiger form \cite{Hutson:2007}:
\begin{equation}
    \delta(E)=\delta_{bg}+\tan^{-1}[\frac{\Gamma_E}{2(E_{res}-E)}],
\end{equation}
where $\delta_{bg}$ is a slowly varying background, $\Gamma_E$ is the resonance width in energy, and $E_{res}$ is the resonance energy. Across a resonance, as $E_{res}-E$ changes sign, and the phase shift thus changes sharply by $\pi$. In the more general inelastic scattering case, multiple channels should be considered and the total phase shift of all channels $\sum_{n}\delta_n(E)$ has to follow the above Breit-Wiger form \cite{Hazi:1979}.

While in our model, multi-channel photoemission is assumed and the EWS time delay can be constructed from a weighted sum \cite{Fanciulli:2018}:
\begin{equation}
    \tau_{EWS}\approx\frac{\sum_{q}\hbar\frac{d \delta^q_{fi}}{d E} |M^q_{fi}|^2}{\sum_{q}|M^q_{fi}|^2}\propto\frac{d \sin(\phi_s)}{dE}\propto \frac{d P}{d E},
\end{equation}
where $\delta_{fi}=\angle M_{fi}$ is the phase of the photoemission matrix element. In the electronic structure of solids, a band is by definition a bound state where such a resonance should occur. Therefore, our observation of a sign reversal of spin polarization across an intensity maximum suggests that, dispersive bands with multiple interfering photoemission channels, in analogy to a Feshbach resonance, provides an aggregate phase shift of $\pi$, and thus a sign reversal of the spin polarization.

\section*{Acknowledgement}

F.G. and J.H.D. acknowledge support from the Swiss National Science Foundation (SNSF) Project No. 200021-200362. M.F. acknowledge support from the Program ERC CZ, project N. LL2314 (TWISTnSHINE).

%------------------------Bibliographie
\bibliographystyle{apsrev4-1}
\bibliography{References_SOIS}

\end{document}